\documentclass[10pt, preprint2]{aastex}
\newcommand{\eg}{e.g.}

\newcommand{\etal}{\mbox{et al.}}
\renewcommand{\*}{$^*$}
\newcommand{\starcore}{$^*_{\rm core}$}
\newcommand{\rcntr}{\mbox{$r_{20\%}$}}
\newcommand{\tabindent}{\noindent \hangindent=24pt \hangafter=1}

\newcommand{\uv}{\mbox{$u$-$v$}}

\newcommand{\Msol}{\mbox{M$_\sun$}}
\newcommand{\Rsol}{\mbox{R$_\sun$}}
\newcommand{\kms}{\mbox{km s$^{-1}$}}
\newcommand{\muas}{\mbox{$\mu$as}}
\newcommand{\muasyr}{\mbox{$\mu$as~yr$^{-1}$}}

\newcommand{\Ra}[4]{\mbox{${#1}^{\rm h} \; {#2}^{\rm m} \; {#3}\fs{#4} $}}
\newcommand{\dec}[4]{\mbox{${#1}\arcdeg \; {#2}\arcmin \; {#3}\farcs{#4} $}}

\begin{document}

\title{SN1993J VLBI (I): The Center of the Explosion and a Limit on Anisotropic
Expansion}

\author{M. F. Bietenholz and N. Bartel}
\affil{Department of Physics and Astronomy, York University, Toronto, M3J~1P3, Ontario, Canada}

\author{and M. P. Rupen}
\affil{National Radio Astronomy Observatory, Socorro, New Mexico 87801, USA}

\shortauthors{Bietenholz \etal}
\shorttitle{SN1993J VLBI}


\begin{abstract}
Phase-referenced VLBI observations of supernova 1993J at 24 epochs,
from 50 days after shock breakout to the present, allowed us to
determine the coordinates of the explosion center relative to the
quasi-stationary core of the host galaxy M81 with an accuracy of
45~\muas, and to determine the nominal proper motion of the geometric
center of the radio shell with an accuracy of 9~\muasyr.  The
uncertainties correspond to 160~AU for the position and 160~\kms\ for
the proper motion at the distance of the source of 3.63~Mpc.  After
correcting for the expected galactic proper motion of the supernova
around the core of M81 using HI rotation curves, we obtain a peculiar
proper motion of the radio shell center of only $320 \pm 160$~\kms\ to
the south, which limits any possible one-sided expansion of the shell.
We also find that the shell is highly circular, the outer contours in
fact being circular to within 3\%.  Combining our proper motion values
with the degree of circular symmetry, we find that the expansion of
the shockfront from the explosion center is isotropic to within 5.5\%
in the plane of the sky.  This is a more fundamental result on
isotropic expansion than can be derived from the circularity of the
images alone.  The brightness of the radio shell, however, varies
along the ridge and systematically changes with time.  The degree of
isotropy in the expansion of the shockfront contrasts with the
asymmetries and polarization found in optical spectral lines.
Asymmetric density distributions in the ejecta or more likely in the
circumstellar medium, are favored to reconcile the radio and optical
results.  We see no sign of any disk-like density distribution of the
circumstellar material, with the average axis ratio of the radio shell
of SN1993J being less than 1.04.

\end{abstract}

\section{INTRODUCTION}

\begin{figure*}
\epsscale{1.8}
\plotone{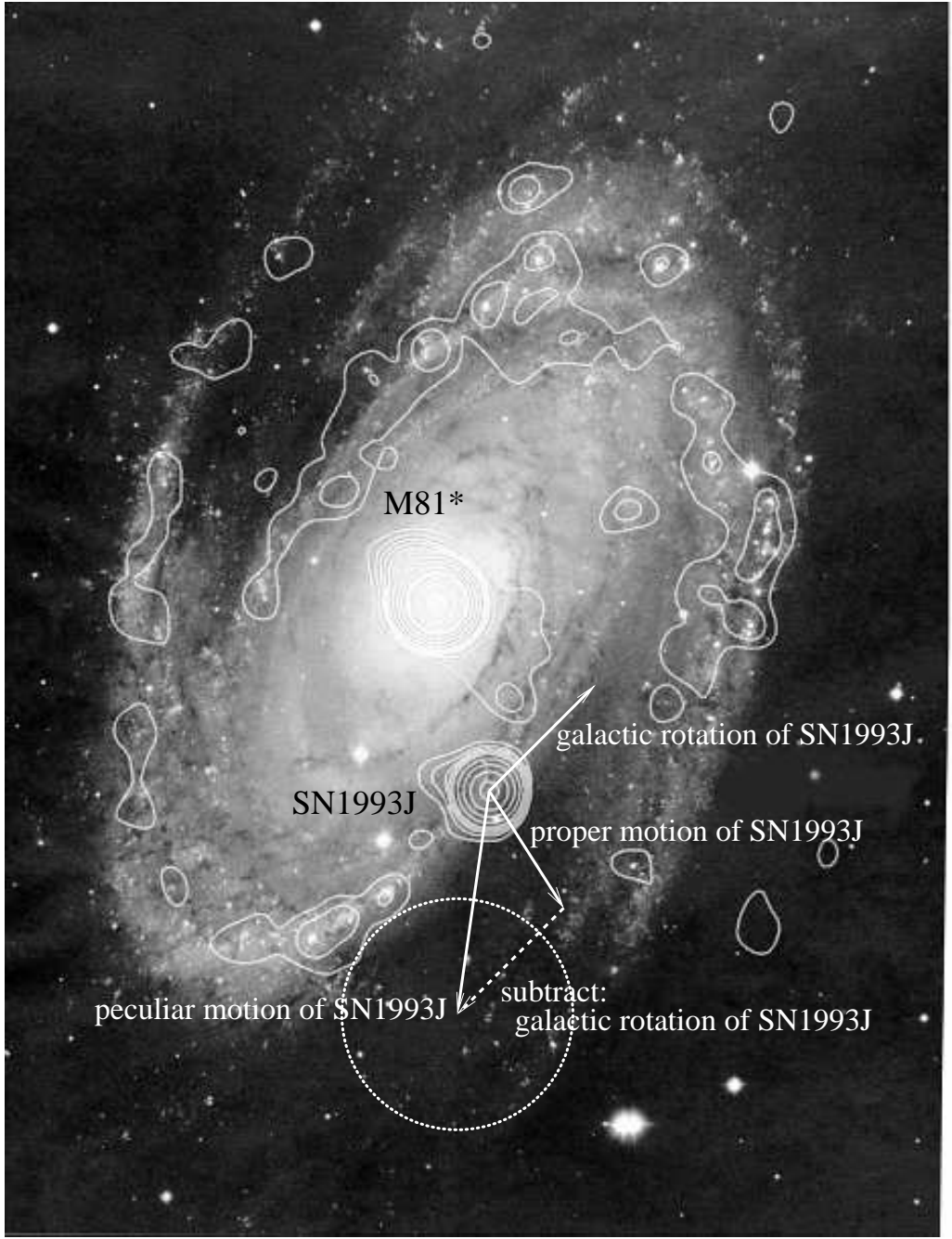}
\figcaption{An optical image of M81 (taken from Sandage, 1961), with
overlaid contours showing the radio emission from M81.  The two
discrete sources indicated are the central radio source, M81\*, and
SN1993J.  The radio data were obtained using the VLA in the D
configuration at 5~GHz on 1997 November 15.  Also shown are the proper
motion of the geometric center of SN1993J, the expected proper
motion of SN1993J due to galactic rotation, and finally the peculiar
proper motion of SN1993J in a co-rotating galactic reference frame,
which are all indicated by the arrows, with the circle representing
the standard error of the peculiar velocity (see text \S~\ref{spropmot}).
\label{fpropmot}}
\end{figure*}

Supernova \objectname[]{SN1993J} was discovered in a spiral arm of
\objectname[]{M81} (\objectname[]{NGC3031}) south south-west of the
galaxy's center by Garcia (Ripero \& Garcia 1993) in the late evening
of 1993 March 28, shortly after shock breakout, which occurred at
$\sim 0$~UT (Wheeler \etal\ 1993) on the same day. It subsequently
became the optically brightest supernova in the northern hemisphere
since SN1954A and also one of the brightest radio supernovae ever
detected (see Figure~\ref{fpropmot} for an optical and radio image of
M81 showing SN1993J near the peak of its radio emission).  At a Cepheid
distance of $3.63\pm0.34$~Mpc (Freedman \etal\ 1994), SN1993J is also
one of the closest extragalactic supernovae ever observed and is
second only to SN1987A as the subject of intense observational and
theoretical supernova studies.  The precursor was identified soon
after the supernova discovery (\eg\ Humphreys \etal\ 1993) and found
to be an approximately K0~I supergiant with a likely mass of $\sim$
17~\Msol\ (Aldering, Humphreys \& Richmond 1994) and a radius $\geq
675$~\Rsol\ or $\geq 3.5$~AU (Clocchiatti \etal\ 1995).

The lightcurve and the spectral properties indicated that SN1993J was
of Type~IIb and was characterized by a low-mass hydrogen outer
layer. H\"oflich, Langer, \& Duschinger (1993) considered a massive
($25 \sim 30$~\Msol) single supergiant that had lost most of its
hydrogen envelope through a strong wind. It is more likely, however,
that the progenitor had the lower mass quoted above and that a close
binary companion had stripped off most of its hydrogen envelope,
leaving it with a residual hydrogen mass in the outer shell of
0.9~\Msol\ or less (Nomoto \etal\ 1993; Shigeyama \etal\ 1994;
Bartunov \etal\ 1994), with most estimates being around 0.2 to
0.4~\Msol\ (\eg\ Podsiadlowski \etal\ 1993; Woosley \etal\ 1994; Houck
\& Fransson 1996).

There were asymmetries in the spectral lines which suggested large
asymmetries (\eg\ Lewis \etal\ 1994; Spyromilio 1994) or clumps
(Matheson \etal\ 2000a, b) in the supernova emission line regions. The
asymmetry in at least one line, however, was found to be possibly due
to line blending (Houck \& Fransson 1996) which weakens somewhat the
conclusions drawn from the asymmetries in other lines. A stronger
indication for asymmetry came from the detection of significant
time-variable optical polarization (Trammell \etal\ 1993; Tran \etal\
1997), and may point to the need for a refinement of the explosion
models.  Most models are symmetric. They have the general
characteristics of an imploding core with the creation of a neutron
star (or possibly a black hole) in its gravitational center, and the
subsequent formation of a shockfront that travels outward
isotropically through the different layers of the progenitor and then
breaks out through the surface. Asymmetric explosions may however
occur if, for instance, the density distribution is not
isotropic. Also, Burrows, Hayes, \& Fryxell (1995) have found through
their modeling of massive star explosions that shortly after the
bounce the expansion has fundamentally anisotropic components and
develops fingers with speeds twice the average expansion
speed. Indeed, pulsars are known to have relatively large proper
motions and a scale height of 400~pc in the $z$-direction away from
the galactic plane (Lyne \& Graham-Smith 1990).  This height is about
five- to sevenfold larger than that of O-B stars (O'Connell 1958) and
supernova remnants (\eg\ Henning \& Wendker 1975) and gives evidence
for asymmetric explosions where the neutron star received a kick at
the time of formation.

Thus an anisotropic velocity pattern may already be present when the
shock front breaks through the surface.  Further anisotropy could
develop when the ejecta start to interact with a circumstellar medium
(CSM) that itself may possess density anisotropies (Blondin,
Lundqvist, \& Chevalier 1996). These latter could be caused by the
particular mass-loss pattern of the progenitor, perhaps with a
predominance in the equatorial plane or perhaps with a bipolar outflow
along the polar axis.  In the binary scenario they could also be
caused by a companion which, in stripping off a sizable quantity of
the envelope in the millions of years before the explosion, would have
substantially distorted the mass-loss density distribution.

To explain the observed polarization and its evolution with time,
H\"oflich (1995) studied three different asymmetric geometries for the
supernova: a) an overall ellipsoid for both inner and outer regions of
the supernova, produced, \eg\ through the differential rotation of the
progenitor; b) an ellipsoidal inner region with a spherical shell,
possibly produced by an aspherical explosion; and c) an overall
spherical geometry with an off-center point source illuminating the
envelope which roughly represents a binary system, or an aspherical
$^{56}$Ni distribution in the ejecta.  H\"oflich \etal\ (1996) favored
model a), an oblate geometry with a ratio between the maximum and
minimum axes or axis ratio of 1.7 seen edge-on, or model c), with
the point source far inside the outer envelope. On the basis of other
polarization data Tran \etal\ (1997) dismissed model a), and instead
favored models b) and c), and in addition a model where the ejecta
interacted with clumpy and anisotropically distributed material in the
CSM.

Radio emission from SN1993J was discovered a few days after shock
breakout (Pooley \& Green 1993; Weiler \etal\ 1993; Phillips \&
Kulkarni 1993) and reached a peak of $\sim$120~mJy at 8.4~GHz.  The
radio lightcurve has been successfully modeled in terms of the
interaction of the expanding shockfront with the CSM (van Dyk \etal\
1994; Fransson \& Bj\"ornsson 1998).  The relatively high flux density,
combined with M81's short distance and high declination,
made SN1993J an ideal target for investigations with the
technique of VLBI, and guaranteed that it would become the best
studied radio supernova ever.

Early VLBI observations resulted in size measurements (Bartel \etal\
1993; Marcaide \etal\ 1993, 1994), determinations of the angular
expansion velocity (Bartel \etal\ 1994; Marcaide \etal\ 1995a) and a
5\% bound on any deviation from circular symmetry of the shape of the
radio source from 30 to 91 days after shock breakout (Bartel \etal\
1994). To reconcile the apparent differences between the radio and the
optical polarization results, Tran \etal\ (1997) speculated that the
explosion was asymmetric in the inner parts of the supernova but
symmetric and isotropic in the outer envelope where the radio
radiation originates.  However, as we will show, circular symmetry
within, say, 5\% alone does not guarantee isotropic expansion within
the same limit. A measurement from a fixed reference point, for
example the explosion center, is needed to properly analyze the
expansion of the radio shell for any anisotropy. Later, the shell-type
morphology of the supernova was revealed (Bartel, Bietenholz, \& Rupen
1995; Marcaide \etal\ 1995b) and the deceleration of the expansion
determined (Marcaide \etal\ 1997; Rupen \etal\ 1998).  We recently
showed that the supernova was almost freely expanding during the first
$\sim 300$ days, but that the expansion decelerated considerably
thereafter, indicating a growing influence of the CSM on the radio
shell and its evolution.  The brightness distribution of the radio
shell changed systematically over time, reflecting perhaps an
anisotropic pattern in the CSM.  After five years, the supernova has
slowed down to half its original expansion velocity, suggesting that
its evolution is now considerably influenced by the swept-up CSM.
These results were summarized by Bartel \etal\ (2000a).

Here we start a series of papers to report in detail on the results
from our comprehensive VLBI observations between 1993 and 2000. For
almost all of our observations, and in particular for all the
observations at 8.4~GHz which we report on here, we used the technique
of phase-referencing with respect to the compact radio source in the
center of M81.  This technique allowed us to obtain the most unbiased
images, and in addition and at least equally important, relative
position determinations with an accuracy of up to 40~\muas. We
recently found a quasi-stationary part in M81's central radio source
which is most likely the core and the gravitational center of the
galaxy (Bietenholz, Bartel, \& Rupen 2000; B00 hereafter). This
finding enables us to investigate the motion of the supernova in a
more fundamental way than if we did not have such a fixed galactic
reference frame.

In this first paper of the series we determine the position of the
explosion center astrometrically via phase-referencing, and
subsequently the degree of isotropy of the expansion of the shock
front from it.  We describe our observations in \S~\ref{obss} and data
reduction in \S~\ref{datreds}.  In \S~\ref{results} we give our
results with representative images of SN1993J, determine the
circularity of the outer contours of all the images, and give a
limit on anisotropic expansion.  In \S~\ref{discuss} we discuss our
results and in \S~\ref{concss} we give our conclusions.  In the
following papers we will present our results on the growing
deceleration of the expansion (paper II), the sequence of images with
the systematically changing structure (paper III), the determination
of the dynamic distance to SN1993J and its host galaxy M81 (paper IV),
and the spectral properties of the radio shell (paper~V).

\section {OBSERVATIONS \label{obss}}

We observed SN1993J at 24 epochs between 1993 and 2000, and at
frequencies between 1.7 and 23~GHz.  We used a global array of between
11 and 18 telescopes, with a total time of 12 to 18 hours for each
run.  SN1993J's high declination of 69\arcdeg\ enabled essentially
100\% visibility at most telescopes, and as a result, we obtained
dense and nearly circular \uv~coverage. As usual, a hydrogen maser was
used as a time and frequency standard at each telescope. The data were
recorded with the VLBA and either the MKIII or the MKIV VLBI systems
with sampling rates of 128 or 256~Mbits per second. The
characteristics of the observations are given in Table~\ref{antab}.
At most epochs, both right and left circular polarizations were
recorded.

All our 8.4 GHz images were phase-referenced, and for this purpose we
alternated short scans of SN1993J and M81\*, the latter being the
compact source in the center of the galaxy M81 and about 170\arcsec\
away from SN1993J (Bietenholz \etal, 1996; B00).  We used a cycle time
of $\sim 3$~min in which SN1993J was observed for 120s and M81\* for
70s, except during the first few months in 1993 when we used a longer
cycle time.  In addition, the sources OQ208 and 0954+658 were observed
occasionally during each observing session as fringe finders and
calibrator sources.  In each session, we recorded data at two to four
frequencies out of a total of six (1.7, 2.3, 5.0, 8.4, 15, and
22~GHz), with 5.0 and 8.4~GHz being the standard frequencies used in
most sessions.  Here we report on results from VLBI observations at
8.4~GHz only.

\section {DATA REDUCTION \label{datreds}}
\subsection{Correlation, Calibration, and Imaging}

The data were correlated with the NRAO VLBA processor in Socorro, New
Mexico, USA.  The analysis was carried out with NRAO's Astronomical
Image Processing System (AIPS) and a modified version of the AIPS task
OMFIT. The initial flux density calibration was done through system
temperature measurements at each telescope.  For each experiment we
first analyzed the data from M81\* by a) fringe-fitting and then b)
iteratively self-calibrating and imaging and/or model-fitting M81\* in
order to determine the complex antenna gains as a function of time.
Phases were derived with a timescale of $\sim 1$~min and amplitudes
with timescales of 1 to 2~h.

The complex antenna gains were then interpolated to the times of the
interleaved supernova scans and applied to the supernova data to
produce phase-referenced images of SN1993J.  Because the two sources
are so close on the sky, this procedure provides an excellent
calibration of the supernova data. Our images are therefore of the
highest quality given the relatively low radio brightness of the
supernova.  In fact, at later epochs, the source is too weak to be
imaged at all without phase-referencing.  In general we thus avoided
the need to self-calibrate the supernova data with an arbitrary
starting model, which can bias the final image especially if the
source is weak.  In case of the first ten epochs, however, the
supernova was still relatively strong and the signal to noise ratio
was sufficient to allow further improvement of the image quality.  We
achieved this improvement by additionally self-calibrating the
antenna-based phases of the supernova data for the first ten epochs,
using as a model the corresponding phase-referenced image of the
supernova.

The images were deconvolved using the CLEAN algorithm. The CLEAN
components were then convolved with a circular Gaussian restoring beam
whose diameter was chosen to be approximately equal to or larger than
the maximum axis of the elliptical Gaussian fit to the inner portion
of the ``dirty beam'' from the uniformly weighted \uv~data. Finally
the residuals from the deconvolution process were added to the CLEAN
components.
Apart from improving the quality of our images for the early epochs
and allowing images to be made at all for the later epochs,
phase-referenced imaging also provided astrometric information
relative to our reference source.  SN1993J is therefore not located at
the phase center of our images as it would be if self-calibration were
used exclusively, but offset somewhat.  Even for the epochs where
self-calibration was used to improve the image quality, the position
of the supernova is known from the strictly phase-referenced image,
and the positions for the shell center that we use later are derived
exclusively from the phase-referenced data before any
self-calibration.  The offset from the phase center reflects a more
accurate position of the center of the supernova with respect to M81\*
than is given by the {\em a priori}\/ coordinates of M81\* and SN1993J
used in the correlator model.  The determination of this offset and
its uncertainty is an important part of our analysis and is
discussed in detail below.

\subsection{Model-fitting \label{smodfit}}

In addition to imaging we used model-fitting. As a model, we used the
two-dimensional projection of a three-dimensional spherical shell of
uniform volume emissivity, which we fit by weighted least squares to
the calibrated \uv~data.  From these model-fits, we derived the
shell's center coordinates, $x_i$ and $y_i$, with respect to the
coordinates of the phase center of the supernova for each epoch, $i$.
These center coordinates, as stated above, were derived prior to any
self-calibration of the supernova data.  Since all group delays, phase
delays, and phase delay rates used in the data analysis are residuals
with respect to the corresponding {\em a priori}\/ delays and rates in
the correlator software model (CALC~6.0 till 1995 Apr.~10; CALC~8.1
till 1999 Nov.~18, and CALC~9.1 thereafter), the phase center simply
corresponds to the nominal coordinates of the supernova used in the
correlator model, for which all residual non-structure delays would be
zero.

We also fit the radius of the shell at each epoch.  For the last
several epochs, where the supernova was sufficiently resolved for the
shell thickness to be estimated with useful accuracy, we further
estimated the outer radius, $r_{\rm out}$, and inner radius, $r_{\rm
in}$, separately.  The mean of the ratio $r_{\rm out}/r_{\rm in}$ for
these epochs was $1.27 \pm 0.02$, and was constant within the
uncertainties (Bartel \etal\ 2000a). Then, for all epochs, we fit
$r_{\rm out}$ only with $r_{\rm in}$ fixed at $r_{\rm
out}/1.25$~\footnote{We use the value of 1.25 as a rounded, earlier
value of the average ratio $r_{\rm out}/r_{\rm in}$.  This small
difference does not significantly change our results.}.  It is from
this last consistent set of fits that the values of $x_i$ and $y_i$
used in this paper were determined.  We will discuss the radii $r_{\rm
out}$ and $r_{\rm in}$ in detail in a future paper.

\section{RESULTS\label{results}}
\subsection{The Degree of Circularity in the Images \label{imags}}

\begin{figure*}
\epsscale{1.0}
\plotone{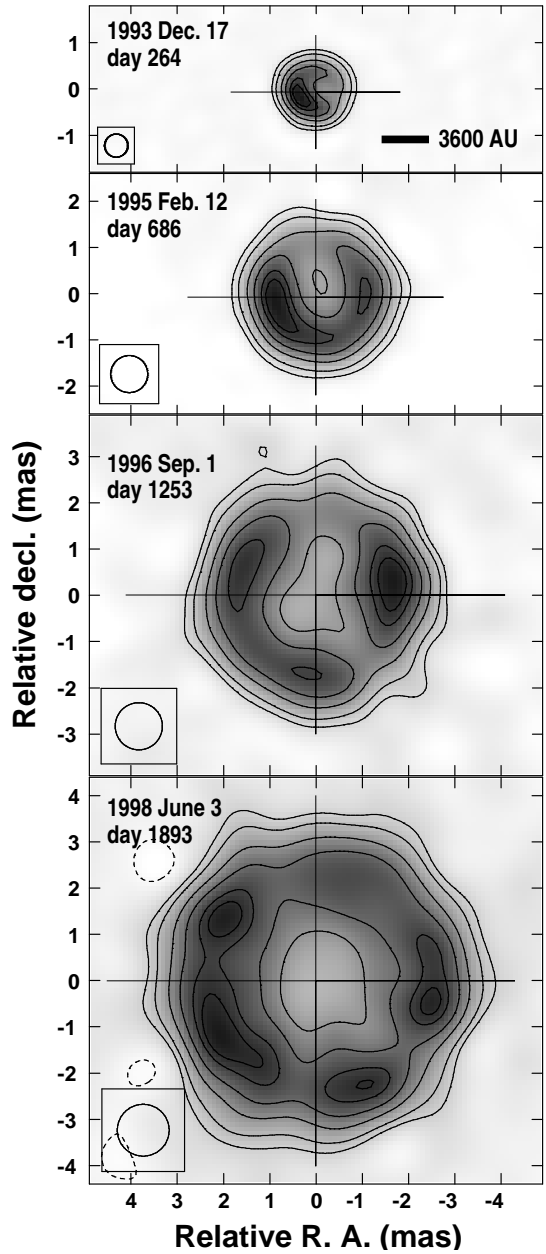} \figcaption{Four representative
images of SN1993J at 8.4~GHz, taken on 1993 December 17, 1995 February
12, 1996 September 1, and 1998 June 3.  The Gaussian restoring beam,
shown in the lower left in each panel, had a FWHM of 0.50, 0.80, 1.02
and 1.12~mas, respectively.  The origin of each panel is the
respective fit geometric center of the shell and is indicated by the
large cross (see text).  The contours on all four images are drawn at
$-10$, 10, 20, 40, 60, 80, and 90\% of the peak brightness, which was,
from the top, 12.1, 3.7, 1.6 and 0.8~mJ~beam$^{-1}$ respectively.  The
bar in the top image shows the linear scale for a distance of 3.63~Mpc.
\label{snmaps}}
\end{figure*}

In Figure~\ref{snmaps} we display four representative 8.4~GHz images
of the supernova at 264~d, 686~d, 1253~d, and 1893~d after shock
breakout. For the origin of the coordinate systems in this figure, we
choose the position of the geometric center of the fit radio shell
(i.e.\ offset at $x_i, \; y_i$ from the phase center position).  For a
full sequence of images up to 2000 February, see Bietenholz \etal\
(2001); see an earlier version until June 1998 in Bartel \etal\
(2000a). A complete sequence of images at 8.4 and 5.0~GHz will be
presented in a future paper.

\begin{figure*}
\epsscale{2.2}
\plotone{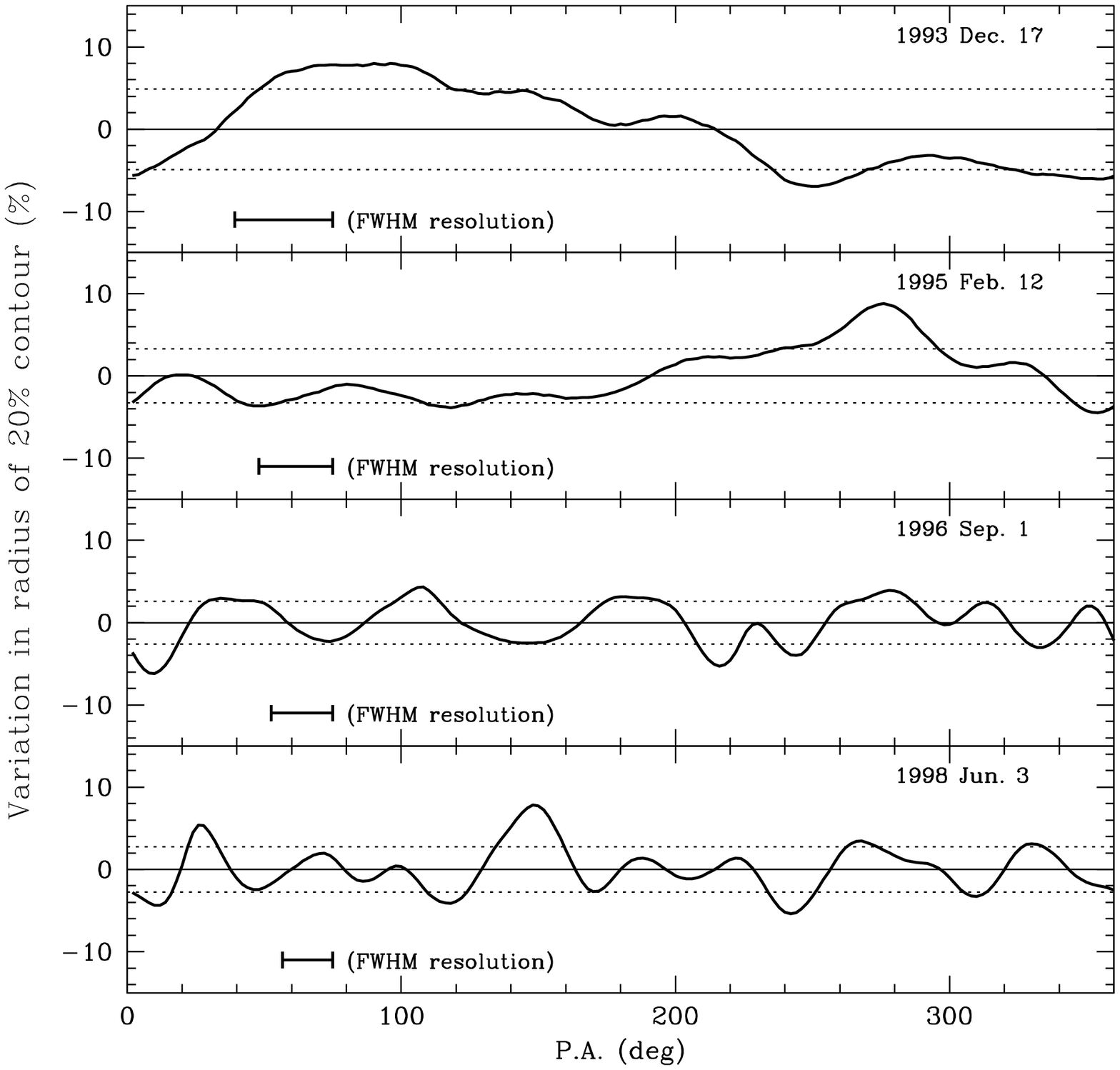}
\figcaption{Plots of the percentage variation of the radius of the
20\% contour of the supernova images (\rcntr) {\em vs.}\/ position
angle (p.a.) for the same representative epochs shown in
Figure~\ref{snmaps}.  The radius and the p.a.\ are given with respect
to the geometric center of the fit shell model (crosses in
Fig.~\ref{snmaps}). The thick line shows the percentage variation of
\rcntr\ around its mean value.  The dotted horizontal lines indicate
the rms of \rcntr, which was, from the top, 4.9\%, 2.3\%, 2.6\% and
2.7\%, respectively.  The angle subtended by the FWHM resolution at
the average value of \rcntr\ is also indicated .
\label{radcnplt}}
\end{figure*}

Although the brightness is modulated along the ridge and is therefore
rotationally asymmetric, the shape of the supernova at the lower
contours is quite circular. We define \rcntr\ to be the radius of the
20\% contour around the center coordinates, $x_i$ and $y_i$, of the
fit shell.  The variation of \rcntr\ with azimuth for a particular
epoch gives a measure of the departures from circular symmetry.  In
Figure~\ref{radcnplt} we show plots of the variation of \rcntr\ about
its mean value {\em vs.}\ the position angle, p.a., for the same
epochs as are shown in Fig.~\ref{snmaps}.

The rms variation of \rcntr\ is less than 5\% for any epoch after
1994.  Taking the average over all the epochs from December~1993
onwards, when the supernova had become sufficiently resolved for a
useful determination of departures from circular symmetry, we find
that the rms deviations from circular symmetry are 3.8\%.  This value
is likely an upper limit to the true deviations from circular
symmetry, since a fraction of the observed deviations must be due to
noise and CLEAN instabilities.  In particular for the latest epochs,
where the dynamic range is relatively low, one would expect a
substantial contribution to the rms in \rcntr\ from noise.  In the
case of small intrinsic variations in \rcntr\ one would expect that
the angular scale of the variation in \rcntr\ would be similar to the
resolution, as is in fact observed (Fig.~\ref{radcnplt}).  If we
average the deviations over 30\arcdeg\ in azimuth, we obtain a
slightly lower figure of 3.1\%, although a substantial fraction of
this value is likely still due to noise.  The real deviations from
circularity not due to the noise or deconvolution are almost certainly
$< 3$\%.

A different measure of circularity is the ellipticity, which is more
sensitive to bimodal distortions.  We determined the ellipticity by
fitting an ellipse to values of \rcntr\ by least squares.  We find
that the maximum observed axis ratio (ratio of the maximum to the
minimum axis) is 1.07.
The (pseudo-vector) average axis ratio is $1.014 \pm 0.008$, at a
p.a.\ of $72\arcdeg \pm 17\arcdeg$.  We consider this ratio not
significantly different from the circular axis ratio of 1.0.  The
slight difference may be due to distortions of the 20\% contour caused
by the eastern and western hot-spots, visible from 1995 on.  We can
accordingly place a formal $3\sigma$ limit on the average axis ratio of
1.04.

\begin{figure}
\epsscale{1.0}
\plotone{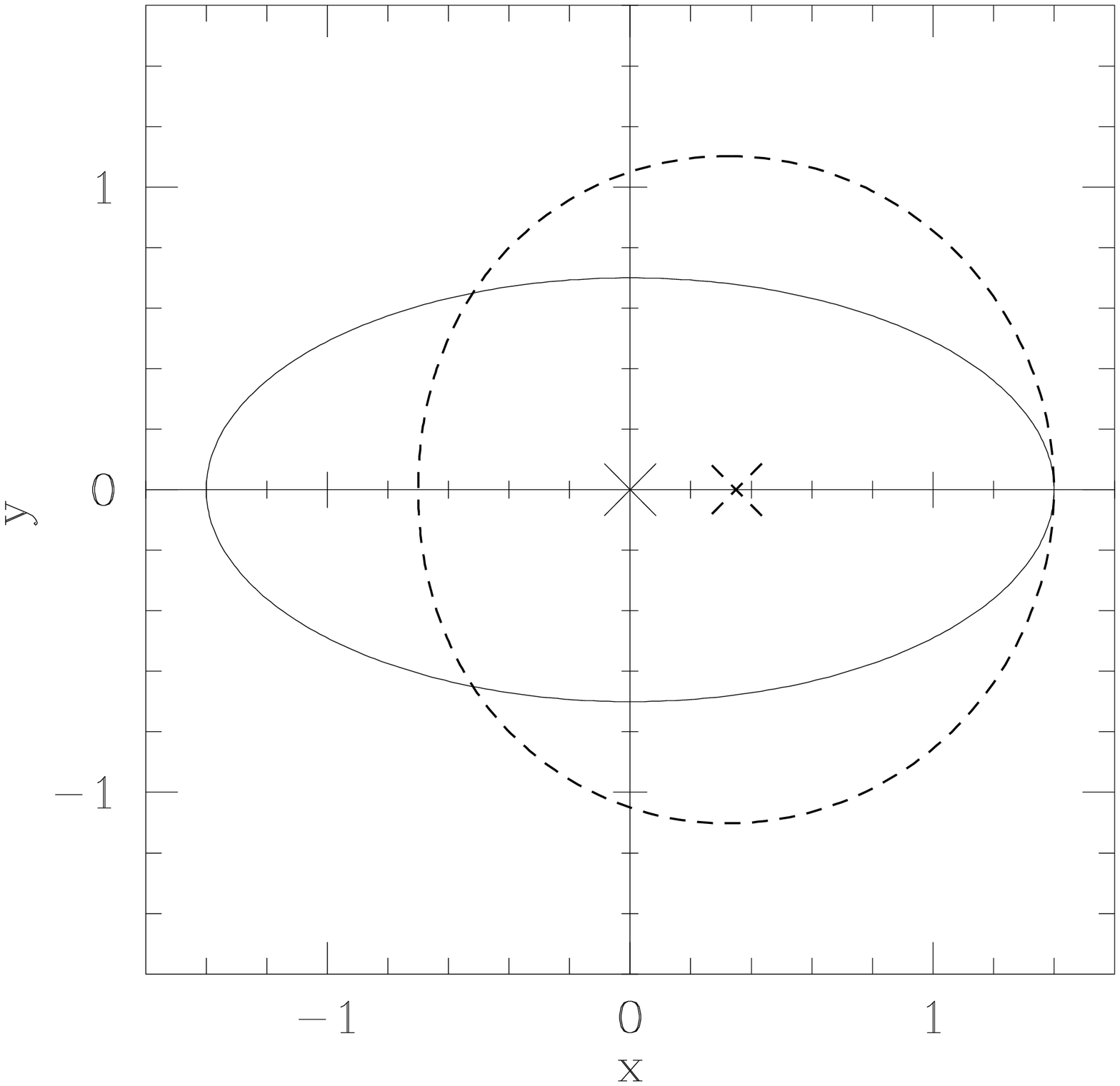}
\figcaption{A sketch showing two different geometries for anisotropic
expansion of a supernova.  The explosion center is at the origin.  In
both cases, the ratio of the maximum to minimum expansion velocity is 2.
The solid line shows a supernova undergoing bipolar expansion, with
the expansion velocities along the $x$ axis being twice those along
the $y$ axis.  In this case, the geometric center of the supernova,
shown by the solid cross, remains at the origin.  The dashed line shows
a supernova undergoing one-sided expansion, with the expansion
velocity at a particular p.a., $\theta$, being proportional to $(1 +
\cos{\theta}\,/3)$, so that the expansion velocity along the
positive $x$ axis is twice that along the negative $x$ axis.  In this
case, the geometric center, shown by the dashed cross, shifts towards
the side of higher velocities.
\label{sngeom}}
\end{figure}

However, these bounds only partly constrain the degree of anisotropic
expansion that the supernova may have undergone. For instance,
differences between expansion velocities along different cuts through
the supernova center would manifest themselves to first order in an
elliptical shape of the supernova. Differences along the same cut in
opposite directions, however, would shift the geometric center of the
supernova in the direction of faster flow, but could leave the shape
of the supernova relatively unchanged. In Figure~\ref{sngeom}, the
solid line illustrates how a supernova which has larger velocities
along the $x$ axis develops an approximately elliptical structure, in
this case with an axis ratio of 2. This ratio also quantifies the
degree of anisotropy of the expansion: the ratio of the maximum to the
minimum expansion velocity is also 2.  The dashed line illustrates a
supernova with the same ratio of the maximum to the minimum expansion
velocity of 2, but where the expansion velocity does not have the same
magnitude in opposite directions, i.e.\ the expansion along the
positive $x$ axis is faster than in the opposite direction.  The
supernova remains circular, but the geometric center of the structure
shifted in the direction of maximum expansion.  In other words, by
measuring only deviations from a circular structure, bipolar and
multipolar expansion are well constrained, but one-sided expansion is
less so. In order to constrain the latter, the center of explosion has
to be determined, and the geometric center coordinates for each epoch,
$x_i$ and $y_i$, have to be plotted with respect to the explosion
center in a fixed reference frame.

\subsection {The Core in M81\*\ as a Reference Point \label{m81refpt}}

As described in \S\ref{smodfit} above, the geometric center
coordinates of the supernova shell, $x_i$ and $y_i$, are determined
relative to the position of the phase center of the corresponding
supernova image for each of our epochs.  The position of the phase
center, however, is subject to variation from epoch to epoch since it
corresponds only approximately to the ``absolute'' position on the
sky.  Errors in the tropospheric and ionospheric delays used at the
correlation stage, as well as small errors in UT1, polar motion, and
antenna coordinates in the correlator model, all contribute to the
positional changes from epoch to epoch, and therefore limit the
accuracy of the phase center position.

Since we phase-referenced to M81\*, the nearby central radio source of
the galaxy, the position of the phase center of the supernova is a
differential position, namely with respect to the position of the
phase center of M81\*, which is essentially the position of the
brightness peak of M81\*. The uncertainty of the differential position
of the supernova phase center is due to only the differential parts of
the above errors in the correlator model, which, over only $\sim
170\arcsec$ on the sky, are much smaller.  In total, the errors in the
correlator model contribute not more than 10~\muas\ to the uncertainty
of a differential position determination of any fiducial point in the
brightness distribution of SN1993J (see Bartel \etal\ 1986, for an
estimate of such errors for the close pair of quasars 3C345 and
NRAO512 at the same frequency).

The brightness peak of M81\*, however, is not a stable reference
point.  As was shown by B00, it is part of a one-sided jet whose
position, size, and orientation are variable on a time scale of
months. The most stable point in the brightness distribution of M81\*\
lies near one end of the brightness distribution, on average
0.3~mas southwest of the brightness peak. B00 identified this point
with the core of M81\*, and therefore we will call it M81\starcore. It
is likely associated with the gravitational center of the galaxy and
can serve as an almost ideal reference point for monitoring the slight
changes in position of the geometric center of the supernova shell.

In fact, the determination of M81\starcore's position is not
independent of the determinations of SN1993J's position: B00
determined the most stable point in the brightness distribution of
M81\* as precisely that one which minimized the variation in the
differential position of the geometric center of SN1993J over all
epochs.  Thus any determination of the changes in the position of
SN1993J with respect to that of M81\starcore\ with our data must
simultaneously involve a determination of the location of
M81\starcore\ itself.  Accordingly, we update the analysis of B00. We
solved simultaneously and by least squares for the following
parameters:

\begin{trivlist}
\item{1.} The displacement of M81\starcore\ with respect to the
brightness peak of M81\*. This displacement is taken to be along the
direction of elongation of M81\* and is measured in units of the
full-width at half-maximum (FWHM) of an elliptical Gaussian fit to the
brightness distribution of M81\* (see B00).

\item{2.} The position of the explosion center, ie., the position of the
geometric center of the radio shell of SN1993J at the time of
explosion.

\item{3.} The proper motion of the geometric center of the radio shell
of SN1993J.

\end{trivlist}

\noindent The input values to this calculation were:

\begin{trivlist}

\item{1.}  The offsets of the geometric center of SN1993J from the
phase center at each epoch, $x_i, \; y_i$.  We determined the total
error of $x_i$ and $y_i$ by adding in quadrature a) the aforementioned
systematic errors of 10~\muas, b) systematic errors induced through
mapping and model-fitting and estimated by monitoring the change of the
center coordinate solutions when only right-circularly or
left-circularly polarized data were used for the fit, and c)
statistical errors from the model-fit. We obtained a rounded total
error of 40~\muas\ for each epoch.

\item{2.}  The times since explosion for each epoch, $t_i$ (no
uncertainties assumed).

\item{3.} The size and orientation of the central elliptical Gaussian
of M81\* at each epoch.  The uncertainties are taken from B00, namely
30~\muas\ in the FWHM and 3\arcdeg\ in the p.a.

\end{trivlist}

\subsection {The Center of Explosion and the Proper Motion of the Supernova \label{spropmot}}

We now give the results from the least squares fit discussed above.
The standard errors for each of the fit parameters are the sum in
quadrature of the statistical uncertainties in the fit, and the rms
scatter obtained from a Monte-Carlo simulation where we varied the
input values randomly according to their uncertainties as given in the
previous section.  We note that the rms of the residuals in position
for the center of SN1993J is 40~\muas, which is consistent with the
value we used for the positional uncertainty.

First we obtained a value for the displacement of M81\starcore\
with respect to the brightness peak of M81\*\ of $0.55 \pm 0.20$~FWHM,
which is consistent with, but more accurate than, the earlier value of
B00.  On average, this value corresponds to $\sim 0.3$~mas, with an
uncertainty at each epoch $\sim 0.1$~mas.

This determination of the position of M81\starcore\ is supported by
preliminary results from our observations of the pair of sources
M81\* and SN1993J at other frequencies, where we found that radio emission
at higher frequencies tends to emanate closer to our core position
(Ebbers \etal\ 1998). We expect to obtain a more accurate value of the
position of M81\starcore\ when the multi-frequency part of our
phase-referencing observations is finished.  Spectral index maps of
M81\* could then be made and the optically thick part of the
brightness distribution of M81\* used for another identification of
M81\starcore.

What is the ``absolute'' position of M81\starcore?  Since the
structure of M81\* is variable from epoch to epoch, the position of
its phase center will be slightly variable from epoch to epoch with
respect to the assumed fixed position of M81\starcore.  We chose the
absolute position of M81\starcore\ such that the average position of
the phase center (or brightness peak) is coincident with the position
of M81\* used in the correlator model, which is tied to the
ICRF\footnote{ICRF: International Celestial Reference Frame}.  We give
this position of M81\* and the position of M81\starcore\ in
Table~\ref{coordparam}.

Second we obtained a value for the coordinates of the explosion
center, $\alpha_{\rm explosion}$ and $\delta_{\rm explosion}$, also
listed in Table~\ref{coordparam}.  It has a standard error of
45~\muas\ which corresponds to $\sim 160$~AU at the distance of M81.

\begin{figure*}
\epsscale{2.2}
\plotone{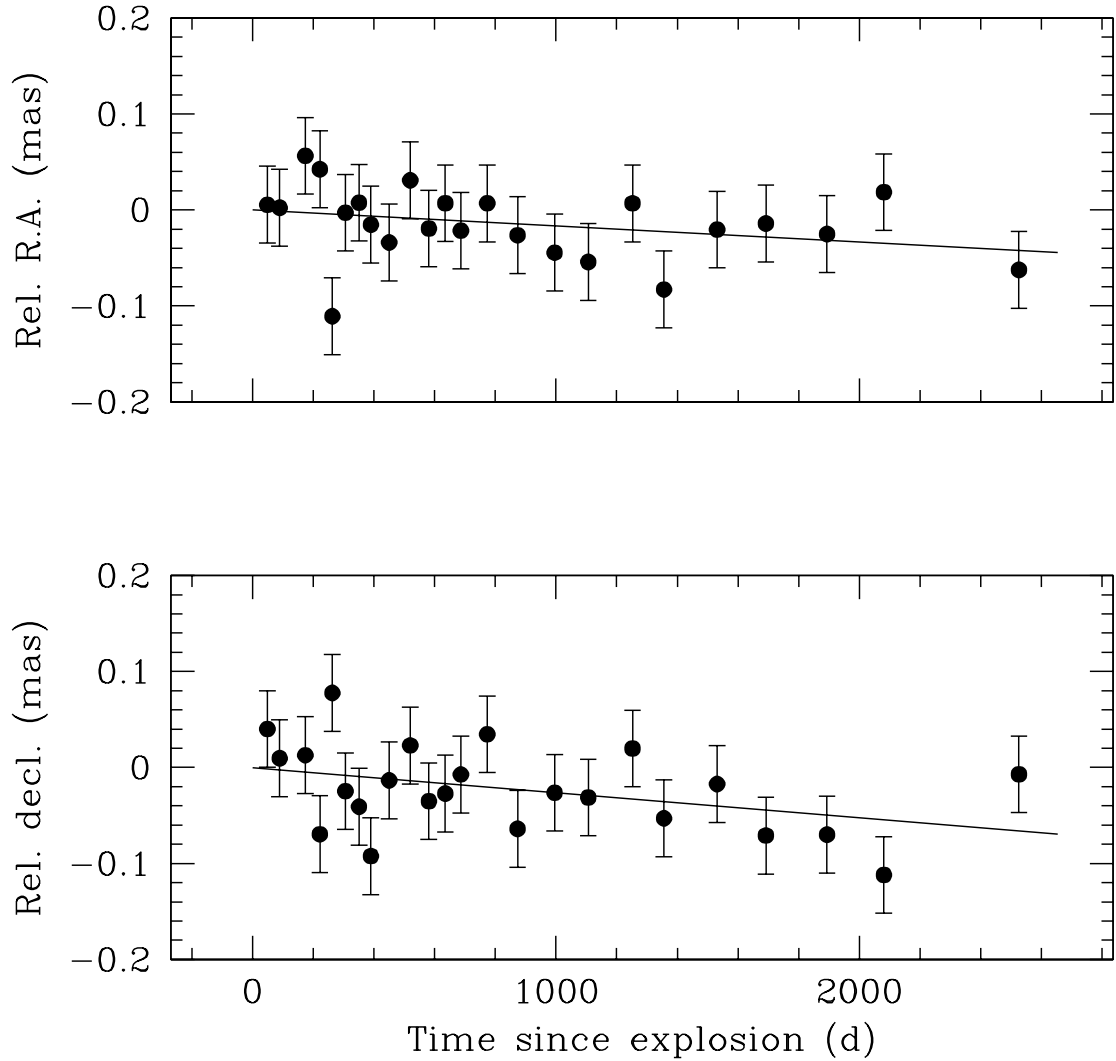}
\figcaption{A plot of $\Delta\alpha_i$ and $\Delta\delta_i$, the
coordinates of the geometric center of the shell of SN1993J
($\alpha_{\rm SNcenter}^i, \; \delta_{\rm SNcenter}^i$) minus those of
the explosion center ($\alpha_{\rm explosion}, \; \delta_{\rm
explosion}$) as a function of time (see Tables~\ref{coordparam},
\ref{cencoord}).  The least squares fit to the proper motion of the
center of the shell is indicated by the straight lines.
\label{sncentr}}
\end{figure*}

Finally we obtained a value for the proper motion of the geometric
center of SN1993J, again listed in Table~\ref{coordparam}.  We list
the coordinates of the geometric center of SN1993J, $\alpha_{\rm
SNcenter}^i$ and $\delta_{\rm SNcenter}^i$, relative to the explosion
center in Table~\ref{cencoord} and plot them in
Figure~\ref{sncentr}. The fit proper motion is displayed as a straight
line in each of the panels.

For a distance to M81 of 3.63~Mpc the proper motion estimate
corresponds to projected velocity components of $-105 \pm 160$~\kms\
in R.A. and $-165 \pm 160$~\kms\ in decl.  We expect galactic proper
motion of the supernova around M81\*, since the supernova most likely
takes part in the galactic rotation.  The expected velocity can be
estimated from the HI rotation curve (Adler \& West\-pfahl 1996; See
also Rots \& Shane 1975) to be $-150 \pm 30$~\kms\ in R.A.
and $+150 \pm 30$~\kms\ in decl., where the uncertainty is also
estimated from the rotation curve and includes a contribution from
possible non-circular motions.  Subtracting the galactic proper motion
from the measured proper motion gives a nominal value for the peculiar
proper motion of the geometric center in the co-rotating galactic
reference frame of $45 \pm 160$~\kms\ in R.A. and $-315 \pm 160$~\kms\
in decl.\ or $320 \pm 160$~\kms\ at p.a.\ $-172 \pm 30$\arcdeg.  These
values and their standard errors are listed in Table~\ref{coordparam}.
We illustrate the various proper motions discussed above on the
optical/radio image of the galaxy M81 in Figure~\ref{fpropmot}.

\subsection {Error Analysis\label{errors}}

How much are our results on the center of explosion and the proper
motion of the geometric center of the supernova dependent on structure
changes of the phase-reference source M81\* and the supernova radio
shell itself? The effects of structure changes of the phase-reference
source were largely compensated for by determining the position of
M81\starcore, rather than using the variable position of the 
brightness peak.  However, since the parameterization of M81\*\ is not
unique, and our determination of the location of M81\starcore\ within
the brightness distribution is uncertain at a level of $\sim
100$~\muas, a small residual motion of M81\starcore\ with respect to
the true location of the black hole is not excluded.  Our
uncertainties include this possibility.  Such a motion would be
expected to be along the average long axis of M81\*.  The fact that
our proper motion for the center of SN1993J w.r.t.\ M81\starcore\ is
roughly along this axis suggests this may indeed be the case, so it is
likely that our peculiar velocity represents in fact an upper limit to
the true peculiar velocity of the center of SN1993J.

To investigate the effects of the structure changes of the supernova
radio shell, we compared the geometric center determined by
model-fitting the \uv~data to the center of the 20\% contour of our
images.  We did not find any significant bias: the vector mean and rms
of the difference over all images after 1993~November were only 11 and
37~\muas, respectively.  However, it is likely that both the model-fit
and the 20\% contour are biased in a similar way.  To check this
possibility, we computed simulations in which we artificially
modulated the brightness distribution of a projected spherical shell.
We found that a simple, sinusoidal modulation in azimuth indeed
somewhat biases the center coordinates of a fit symmetric model
towards the brighter side of the image and the radius towards smaller
values.  However, these biases are smaller than our uncertainties for
any realistic degree of such modulation, and they become yet smaller for
more complex modulations.

In fact, when we exclude the data from the first 520 days which, if
any, are most likely to be biased, we obtain coordinates of the center
of explosion and a proper motion different by less than our stated
standard errors.  We conclude that there is no evidence for any
significant systematic effects on the coordinates of the explosion
center and the proper motion of the geometric center of the supernova
from structure changes of either M81\* or the supernova radio shell.

\subsection{Limit on Anisotropic Expansion}

While the peculiar proper motion of the geometric center of the radio
shell is at the $2\sigma$ level and therefore hardly significant, it
is not in fact expected to be zero. First momentum conservation
dictates that the supernova shell and the putative pulsar must inherit
the momentum of the progenitor.  The shell as a whole would thus be
expected to have a proper motion similar to that of the progenitor.
The peculiar motion of the progenitor, however, is almost certainly
considerably less than 320~\kms: the peculiar velocity dispersion of
F-M supergiants in the solar neighborhood is $<15$~\kms\ (Binney \&
Merrifield 1998).  Second anisotropic expansion of the supernova
shell would lead to a proper motion of its geometric center.  Since
the contribution of the progenitor's peculiar proper motion is likely
insignificant, the proper motion of the center of the supernova shell
can thus give us information about possible anisotropic expansion.

In particular, we can constrain the degree of anisotropic expansion by
combining the peculiar proper motion of the geometric center in the
co-rotating galactic reference frame with the bound on any deviation
from circular symmetry of the supernova radio shell in the plane of
the sky. The $1\sigma$ bound on the proper motion of the center of the
shell is 28~\muasyr, so after five years, at the time of the
observations of the radio shell in the last panel of
Figure~\ref{snmaps}, the $1\sigma$ bound on the distance traveled by
the geometric center of the supernova is 150~\muas. This distance
limit is 4.6\% of the outer radius of the radio shell. Adding this
value in quadrature to the bound on circularity deviations (over
30\arcdeg\ in azimuth) of 3.1\% gives a bound on anisotropic expansion
of the radio shell of 5.5\%.

\section{DISCUSSION \label{discuss}}

Twenty four epochs of observations of SN1993J, over nearly seven years
from the time of explosion to the present, phase-referenced to the
almost ideal reference point of the core of the host galaxy, have
significantly advanced our knowledge of the evolution of supernova
radio shells. In our Galaxy radio shells of supernovae have been
observed at most over about 10\% of the supernova's age.  SN1993J has
been observed over essentially 100\% of its age. In this first of a
series of papers, we determined the position of the explosion center in
the galactic reference frame with an accuracy of $\sim 160$~AU. This is
the reference point with respect to which the expansion of the radio
shell can best be measured in a physically meaningful way, from
the first tens of days after the explosion to the present and for
years to come. We determined an upper limit on any anisotropic
expansion in the plane of the sky of 5.5\%.
This result is important as a limiting factor in discussions
concerning an asymmetry of the explosion, the recoil velocity of a
central pulsar that is expected to have formed, distortion of the
ejecta velocity pattern by a binary companion, asymmetries in the CSM
caused by anisotropies in the mass-loss to wind-velocity ratio of the
progenitor or by the influence of a binary companion, and in
determining part of the error of the dynamic distance estimate of
SN1993J and its host galaxy. We will discuss each of the aspects in
turn.

\subsection {Constraint on Asymmetry of the Explosion}

Recent modeling of core-collapse supernova explosions (Burrows \etal\
1995) suggests that they have fundamentally heterogeneous and
anisotropic components. During the first several tens of milliseconds
after the bounce when the explosion commences, plumes and fingers can
emerge with velocities twice as large as the velocity of the shock wave
itself. Those authors find that these characteristics of the explosion
process are consistent with the asymmetries found in the infrared line
profiles of SN1987A and the ``shrapnel'' observed in some supernova
remnants, like Cas~A and Vela.  SN1993J provided us with images from
the earliest times after the explosion of any supernova yet, and may
therefore be of particular interest for searching for traces of such
jet-like features from the explosion process. No such anisotropies,
however, have yet been found. If they are as yet unresolved, then they
could perhaps become visible in future observations, when we will have
increased relative resolution due to the continuing expansion of the
supernova.

Asymmetries in the ejecta are indicated by the time-variable optical
polarization (Trammell \etal\ 1993; Tran \etal\ 1997) and perhaps by
the optical spectral lines (\eg\ Lewis \etal\ 1994; Spyromilio 1994;
Matheson \etal\ 2000a, b).  With the addition of the astrometric
result, our current limit on any anisotropic expansion of the outer
rim of the radio shell in the plane of the sky is more stringent than
our previous limit, which was based on the degree of circular symmetry
of the radio shell alone.  If the velocity isotropy we measure for the
shock front in the plane of the sky is a good indication of the
degree of velocity isotropy in space, then H\"oflich's (1995) model a)
of an overall elliptical structure of the supernova produced \eg\
through rotation of the progenitor would have to be dismissed.  The
only way to reconcile the results from the optical with those from the
radio observations in terms of asymmetries in the ejecta is to assume
a model along the lines of H\"oflich's (1995) model b) of an
ellipsoidal inner region with a spherical shell, or model c) of an
overall spherical geometry with an off-center point-like source
represented, \eg, in the binary scenario.

In this context it is of interest to compare the optical with the
radio results for SN1987A, the only other supernova for which both
optical polarization measurements and VLBI images have been
obtained. Based on the optical polarization, several authors (\eg,
Barrett 1988; Cropper \etal\ 1988; Mendez \etal\ 1988; Jeffery 1991)
concluded that the ejecta were characterized by asymmetries of at
least 10\%. The radio images of SN1987A  (Gaensler \etal\ 1997) appear
relatively round, although asymmetries of up to 10\% are not excluded.
The only other imaged radio supernova is SN1986J, and its geometry is
less circular than either SN1993J or SN1987A (Bartel \etal\ 1991).
Furthermore, young supernova remnants in M82 have been imaged, and they
are also less circular (e.g.\ Pedlar \etal\ 1999; Bartel \etal\ 1987).
Different ejecta characteristics and/or circumstellar environments may
have contributed to the asymmetries. Also, the supernova remnants in
M82 were imaged when they were much older than SN1993J was at our last
epoch, so it is possible that asymmetries grow with time and may
become apparent in SN1993J in future observations.

\subsection {Limit on the Recoil Velocity of a Central Pulsar}

The average speed of radio pulsars at birth is $\sim 450$~\kms\ (Lyne
\& Lorimer 1994; Frail, Goss, \& Whiteoak, 1994), which is
significantly higher than that of stars.  The most natural explanation
is that neutron stars receive a kick when they are formed.
Asymmetries in the explosion may cause such kicks (\eg\ Burrows \etal\
1995).  Since the pulsar expected from SN1993J is (as yet) undetected
(Bartel \etal\ 2000b) we cannot directly determine its velocity, but
the assumption of momentum conservation allows us to estimate an upper
limit.  It is unlikely that the pulsar acquires more than, say, 1/4 of
the momentum imparted to the shell.  Then, taking our proper motion of
the shell's geometric center of 320~\kms, a shell mass of 16~\Msol,
and a neutron star mass of 1.4~\Msol, and assuming the progenitor's
momentum to be negligible, we find the pulsar's projected velocity
unlikely to exceed $\sim 1000$~\kms.  This implies that the pulsar is
still within 0.4~mas of the center of the shell.

\subsection {Constraint on Anisotropy in the Density of the CSM}

Anisotropy in the density distribution of the CSM could be responsible
for any non-sphericities the radio shell may display.  This could be
caused, for example, by a supernova exploding near a molecular cloud.
Dohm-Palmer \& Jones (1996) have investigated this case, and find that
the supernova shell remains almost circular in the early stages, but
that the geometric center shifts towards the region of lower density.
Such anisotropy could also be caused by a latitude-dependent
ratio of mass-loss to wind-velocity, $\dot M / w$, leading, \eg, to a
bipolar outflow along the polar axis of the progenitor, or to an
equatorial outflow resulting in a disk-like density distribution.
The influence of a binary companion would likely also cause an
anisotropic CSM.

Blondin, Lundqvist, \& Chevalier (1996) used hydrodynamic modeling to
study how the interaction zone deforms when isotropic ejecta expand
into a CSM with an anisotropic density distribution.  Assuming density
profiles for the ejecta and the CSM of $\rho_{\rm ej} \propto r^{-n}$
and $\rho_{\rm CSM} \propto r^{-s}$ with $n=7$ and $s=2$,  they
studied the axis ratio of the self-similar density wave, or by
implication the radio shell, as a function of a model axisymmetric
about the polar axis and reflection symmetric about the equator with
different ratios of the equatorial and polar densities.
They found that for small polar to equatorial density ratios, the
axial ratio of the supernova shell was only slightly larger than that
expected in the case of purely radial motion, where the axial ratio is
proportional to the density ratio to the power of $1/(n - s)$.

The assumed values for $n$ and $s$ in Blondin \etal's model are
consistent with those for the late evolution of SN1993J (Bartel \etal\
2000a), and their model can therefore be used to place useful limits on
the density distribution of the CSM around SN1993J.  Extrapolating
from Blondin \etal\ we find, for times applicable to those of our
SN1993J measurements, that our upper limit on the axis ratio of 1.04 gives
a limit of the density ratio of approximately 1.2.  This limit
indicates that a spherically symmetric density distribution dominates,
and that any disk-like component of the density distribution is small.
In any case, deformations of the radio shell, whether they are caused
by a disk-like or an irregular density distribution of the CSM, may
need time to develop. The expansion velocity of the shockfront had
already slowed from about $\sim 18,000$~\kms\ at the start of our
observations to $\sim 8,000$~\kms\ by June~1998 (Bartel \etal\ 2000a),
indicating that swept-up material of the CSM already dominates the
evolution of the radio shell. It may be just a matter of time till
significant protrusions and other asymmetries develop. Having
determined the center of explosion and being able to identify that
center in any image at future observing epochs will make us very
sensitive in measuring such asymmetries.

\subsection {The Binary Scenario}

In the likely case that the progenitor of SN1993J had a binary
companion, that companion could perhaps have influenced the density
distribution of the CSM.  Lundqvist (1994) argues that the result of
such influence would be a disk-like density distribution.  We see no
sign of a disk-like density distribution in our images.  Our limit on
the axis ratio can constrain any disk-like density distribution,
provided it extends beyond the radius of our first image or $\sim
500$~AU.  This extent, however, is much larger than the orbital size
of any plausible binary system, and so it is not clear whether a
disk-like density distribution of this size might realistically result
from the influence of the binary companion.

If we postulate a disk of extent $> 500$~AU, then our limit on the
axis ratio above suggests either that the density distribution in fact
resembles a slightly flattend sphere rather than a disk, or that we
are seeing the disk almost face on.  In the latter case, the evolving
azimuthal variations along the ridge of SN1993J might be a reflection of
the azimuthal density distribution and its changes with time.

\subsection {Improving the Accuracy of a Dynamic Distance Estimate}

Finally, the bound on anisotropy in the radio shell velocities is
important for determining the uncertainty of a dynamic distance
estimate for SN1993J and its host galaxy M81. The distance can be
determined directly (Bartel 1984; Bartel \etal\ 1985) by combining
the radial velocity in optical spectra, determined from the blue
edge of the H$\alpha$ absorption trough (Trammell, Hines, \& Wheeler
1993; Lewis \etal\ 1994) with the transverse angular velocity of the
outer boundary of the radio shell.  One component in the error budget
for the distance is the degree to which radial and transverse
velocities are the same. Save for conspiracies, our limit on
anisotropic velocities in the plane of the sky also gives a reasonable
limit on the anisotropic velocities in three dimensions, and
is therefore useful for constraining the error of the distance. A
paper on the distance determination is in preparation.

\section {CONCLUSIONS \label{concss}}

\noindent Here we list a summary of our main conclusions.

\begin{trivlist}

\item{1.} We determined the center of explosion of SN1993J relative to
M81\starcore\ with a standard error of 45~\muas\ or $\sim 160$~AU, only
$\sim 40$ times the estimated radius of the progenitor.  

\item{2.} We could thus accurately determine the location of the
center of explosion in each of the VLBI images.  We can also
accurately determine the location of the explosion center in
images from future phase-referenced observations of SN1993J.

\item{3.} Despite the brightness varying with azimuth along the ridge,
we found no protrusions. The outer contour of the radio shell is
circular to within 3\%.

\item{4.} The geometric center of the radio shell has a nominal proper
motion of $-6 \pm 9$~\muasyr\ in R.A. and $-10 \pm 9$~\muasyr\ in
decl.\ relative to M81\starcore\ which is presumably the black hole
and gravitational center of the galaxy.  For a distance of 3.63~Mpc,
the values correspond to velocity components of $-105 \pm 160$~\kms\
and $-165 \pm 160$~\kms, respectively. After correction for galactic
HI rotation, the peculiar proper motion of the geometric center of the
radio shell relative to M81\starcore\ is $45 \pm 160$~\kms\ in
R.A. and $-215 \pm 160$~\kms\ in decl.

\item{4.} The expansion velocity of the shockfront is isotropic to
within 5.5\% in the plane of the sky.

\item{5.} The circularity of the outer contours of the radio shell and
the degree of isotropic expansion of the shockfront constrain the
large-scale density distribution of the circumstellar medium.  We saw
no sign of a disk-like density distribution of the CSM.  The average
axis ratio for elliptical fits to the 20\% contour of the shell was
$<1.04$.  If a disk-like density distribution was formed by the
purported binary system before the explosion, and if such distribution
extends over 1000's of AU, then we are seeing the disk close to
face-on.

\item{6.}  The swept-up material is likely to have an increasing
influence on the evolution of the radio shell in the years to come.
Significant anisotropic expansion may develop in the future, and any
anisotropy will likely first be detected reliably with the combination
of imaging and astrometry presented in this paper.

\end{trivlist}

\acknowledgements

ACKNOWLEDGMENTS.  N. Bartel thanks the Canadian Institute for
Theoretical Astrophysics (CITA) in Toronto and the Observat\'{o}rio
Nacional in Rio de Janeiro for their hospitality and support during
part of his sabbatical year while this paper was being written.  We
thank V. I. Altunin, A. J. Beasley, W. H. Cannon, J. E. Conway,
D. A. Graham, D. L. Jones, A. Rius, G. Umana, and T. Venturi for help
with several aspects of the project.  OMFIT was written by K. Desai.
We thank NRAO, the European VLBI Network, the NASA/JPL Deep Space
Network (DSN), and Natural Resources Canada for providing support for
the observations. Research at York University was partly supported by
NSERC.  NRAO is operated under license by Associated Universities,
Inc., under cooperative agreement with NSF.  The NASA/JPL DSN is
operated by JPL/Caltech, under contract with NASA.  We have made use
of NASA's Astrophysics Data System Abstract Service.

\clearpage
\onecolumn

\begin{deluxetable}{r@{ }l@{ }r c  c@{ }c@{ }c@{ }c@{ }c@{ }c@{ }c@{ }c@{ }c@{ }c@{ }c@{ }c@{ }c@{ }c@{ }c@{ }c@{ }c@{ }c@{ }c@{ }c@{\hspace{0.3in}} c c c}
\tabletypesize{\footnotesize}
\rotate
\tablecaption{8.4 GHz VLBI Observations of SN1993J\label{antab}}
\tablewidth{570pt}
\tablehead{
%
%
& & & 
 &\multicolumn{20}{c}{ANTENNA\tablenotemark{a}} & \colhead{Total} & \colhead{On-Source} &\colhead{Recording} \\
%
%
\multicolumn{3}{c}{Date}  & \colhead{Freq.}
 &\multicolumn{20}{c}{~}
 &\colhead{time\tablenotemark{b}} & \colhead{time\tablenotemark{c}} & \colhead{Mode\tablenotemark{d}} \\
%
%
 & & & \colhead{(GHz)} & Eb & Wb & Mc & Nt & On & Go & Ro & Aq & Gb &  Y & Br & Fd & Hn & Kp & La & Mk & Nl & Ov & Pt & Sc
 &\colhead{(hr)}  &\colhead{(baseline-hr)} 
}
\startdata
%
 1993&May&17 & 8.4 & X &   &  &   &   &   &   & X &   & X & X & X & X & X &   & X & X & X & X & X & 
   \phn 9.6  & \phn 41   & III-B\\
 1993&Jun&27 & 8.4 &   &   &  &   &   &   &   &   &   &   & X & X & X & X & X & X & X &   & X & X & 
  13.0  & \phn 41    & III-B\\
 1993&Aug&4\tablenotemark{e} 
             & 8.4 &   &   &   & X &   &   &   & X &   &   & X & X &   & X & X &   & X &   & X & X & 
  16.6  & \phn 15    & III-B\\
 1993&Sep&19 & 8.4 &   &   & X &   &   &   & X & X &   & X & X & X & X & X & X & X & X & X & X & X & 
   17.6  &     164   & III-B\\
 1993&Nov& 6 & 8.4 & X &   & X &   &   &   & X &   &   & X & X & X & X & X & X & X & X & X & X & X & 
   18.0  &     202   &III-B\\
 1993&Dec&17 & 8.4 & X &   & X &   &   &X\*&   & X &   & X & X & X & X & X & X & X & X & X & X & X & 
   18.0  &     133   & III-B\\
 1994&Jan&28 & 8.4 & X &   &   &   &   &   &X\*&   &   & X & X & X & X & X & X &   & X & X & X & X & 
   17.6  & \phn 85   & III-B\\
 1994&Mar&15 & 8.4 & X &   &   & X &   &   &X\*&   & X & X & X & X & X & X & X & X & X & X & X & X & 
   18.4  &     273   & III-B\\
 1994&Apr&22 & 8.4 & X &   &   & X &   &X\*&X\*&   & X & X & X & X & X & X & X & X & X & X & X & X & 
   16.9  &     159   & III-B\\
 1994&Jun&22 & 8.4 & X &   &   & X &   &X\*&X\*&   & X & X & X & X & X & X & X & X & X & X & X & X & 
   16.1  &     216   & III-B\\
 1994&Aug&30 & 8.4 & X &   &   & X &   &   &X\*&   &   & X & X & X & X & X & X & X & X & X & X & X & 
   14.8  &     135   & III-B\\
 1994&Oct&31 & 8.4 &   &   & X & X &   &   & X & X &   & X & X & X & X & X & X & X & X & X & X & X & 
   15.1  &     134   & III-B\\
 1994&Dec&23 & 8.4 & X &   &   & X &   & X & X & X & X & X & X & X & X & X & X & X & X &   & X & X & 
   16.1  &     233   & III-B\\
 1995&Feb&12 & 8.4 & X &   & X & X &   & X & X & X & X & X & X & X & X & X & X & X & X & X & X & X & 
   11.8  &     641   & III-B\\
 1995&May&11 & 8.4 &   &   &   &   &   &   &   &   &   & X & X & X & X & X & X & X & X & X & X & X & 
   15.4  & \phn 71   & 128-4-2\\
 1995&Aug&18 & 8.4 &   &   &   &   &   &   &   &   &   & X & X & X & X & X & X & X & X & X & X & X & 
   14.2  & \phn 59   & 128-4-2\\
 1995&Dec&19 & 8.4 &   &   &   &   &   &   & X &   &   & X & X & X & X & X & X & X & X & X & X & X & 
   15.7  &     121   & 128-4-2\\
 1996&Apr& 8 & 8.4 &   &   &   &   &   & X & X &   &   & X &   & X & X & X & X & X & X & X & X & X & 
   \phn 7.4  & 190   & III-B\\
 1996&Sep& 1 & 8.4 & X &   &   &  &   &   &   &   &  & X & X & X & X & X & X & X & X & X & X & X & 
   16.4  &     133   & 128-4-2\\
 1996&Dec&13 & 8.4 & X &   &   &   &   &   &   &   &   & X & X & X & X & X & X & X & X & X & X & X & 
   17.2  &     123   & 256-8-2\\
 1997&Jun& 7 & 8.4 & X &   &   &   &   &   &   &   & X & X & X & X & X & X & X & X & X & X & X & X & 
   17.2  &     114   & 256-8-2\\
 1997&Nov&15 & 8.4 & X &   & X & X &   & X & X &   & X & X & X & X & X & X & X & X & X & X & X & X & 
   13.8  &     266   & 256-8-2\\
 1998&Jun& 3 & 8.4 & X &   & X &   &   &   &   &   &   & X & X & X & X & X & X & X & X & X & X & X & 
   10.9  &     300   & 256-8-2\\
 1998&Dec& 7 & 8.4 & X &   & X & X &   & X & X &   &   & X & X & X & X &   & X & X & X & X & X & X & 
   12.2  &     304   & 256-8-2\\
 2000&Feb&25 & 8.4 & X & X & X & X & X & X & X &   &   & X & X & X & X &   & X & X & X & X & X & X & 
  \phn 9.4  &  418    & 256-8-2\\
\enddata
\tablenotetext{a}{
  Ef= 100m, MPIfR, Effelsberg, Germany;
  Wb= equivalent diameter 94m, Westerbork, the Netherlands;\phn
  Mc=  32m, IdR-CNR, Medicina, Italy;
  Nt=  32m, IdR-CNR, Noto, Italy;
  On=  20m, Onsala Space Observatory, Sweden;
  Go=  70m, NASA-JPL, Goldstone, CA, USA (Asterisk denotes use of the 34m antenna
             at the same station);
  Ro=  70m, NASA-JPL, Robledo, Spain (Asterisk denotes use of the 34m antenna
             at the same station);
  Aq=  46m, ISTS (now CRESTech/York Univ.), Algonquin Park, Ontario, Canada;
  Gb=  43m, NRAO, Green Bank, WV, USA;
  Y = equivalent diameter 130m, NRAO, near Socorro, NM, USA;\phn
  Br=  25m, NRAO, Brewster, WA, USA;
  Fd=  25m, NRAO, Fort Davis, TX, USA;
  Hn=  25m, NRAO, Hancock, NH, USA;
  Kp=  25m, NRAO, Kitt Peak, AZ, USA;
  La=  25m, NRAO, Los Alamos, NM, USA;
  Mk=  25m, NRAO, Mauna Kea, HI, USA;
  Nl=  25m, NRAO, North Liberty, IA, USA;
  Ov=  25m, NRAO, Owens Valley, CA, USA;
  Pt=  25m, NRAO, Pie Town, NM, USA;
  Sc=  25m, NRAO, St. Croix, Virgin Islands, USA.
  }
\tablenotetext{b}{Maximum span in hour angle at any one antenna.}
\tablenotetext{c}{Number of baseline-hours spent on SN1993J, after data calibration and editing.}
\tablenotetext{d}{Recording mode: III-B= Mk III mode B double speed; \\
128-4-2= VLBA format, 128 MHz recorded in 4 baseband channels with 2-bit sampling. \\
256-8-2= VLBA format, 256 MHz recorded in 8 baseband channels with 2-bit sampling.}
\tablenotetext{e}{Data from this epoch were not used in this paper due to the poor 
convergence of the M81\* model-fit}

\end{deluxetable}

\begin{table}
\caption{Summary of the Astrometric Results}\label{coordparam}
\begin{minipage}{5.0in} 
\begin{tabular}{l l l}
\tabletypesize{\footnotesize}

\\
\tableline\tableline
\hangindent=24.pt
                        &\hspace{10pt} R.A. (J2000) &\hspace{10pt} decl.\ (J2000) \\[3pt]
\tableline
%
Position of M81\* (ICRF)\tablenotemark{a} 
                               & \Ra{9}{55}{33}{173103}
                               & \dec{69}{3}{55}{061630} \\[5pt]

Position of M81\starcore\tablenotemark{b}
                               & \Ra{9}{55}{33}{173063}
                               & \dec{69}{3}{55}{061464} \\[5pt]

\parbox[t]{200pt}{\tabindent Position of explosion center\tablenotemark{c}}
                               & \Ra{9}{55}{24}{7747593} (85)
                               & \dec{69}{1}{13}{703188} (45) \\[20pt]
\singlespace
                         &\hspace{23pt} $\mu_{\rm RA}$ &\hspace{23pt} $\mu_{\rm dec}$ \\[3pt]
\tableline
\parbox[t]{200pt}{\tabindent Proper motion of geometric center of radio shell\tablenotemark{d}}
    & $\; -6.1 \pm 9.3$ \muasyr & $\, -9.6 \pm 9.3$ \muasyr \\

\parbox[t]{200pt}{\tabindent Projected velocity of geometric center of radio shell\tablenotemark{e}}
    &     $ -105 \pm 160$ \kms  &     $ -165 \pm 160$ \kms \\
\\
\parbox{200pt}{\tabindent Projected velocity of HI gas at SN position\tablenotemark{f}}
    &     $ -150 \pm \phn 30$ \kms  & \phs $ 150 \pm \phn 30$ \kms \\
\\
\parbox{200pt}{\tabindent Projected velocity of geometric center of SN in the co-rotating 
               galactic reference frame\tablenotemark{g}}
 & \phn\phs $ 45 \pm 160$ \kms  &     $ -315 \pm  160$ \kms \\
\end{tabular}

\tablenotetext{}{All uncertainties are standard errors; all velocities calculated
for a distance of 3.63~Mpc.}

\tablenotetext{a}{This position was used in the correlator model and is tied to the
International Celestial Reference Frame (ICRF) and essentially refers to the
brightness peak of M81\*.}

\tablenotetext{b}{The position was computed by adding the average
offset of M81\starcore\ from the brightness peak, which was 0.3~mas to
the southwest, to the position of M81\*. It was used as a
reference position for the determinations of the coordinates of the
geometric center of the supernova ($\alpha_{\rm SNcenter}^i, \;
\delta_{\rm SNcenter}^i$) for each epoch, $i$.  For more information, see
text \S\S \ref{m81refpt} -- \ref{spropmot}.}

\tablenotetext{c}{The position ($\alpha_{\rm explosion}, \;
\delta_{\rm explosion}$) was determined as the intercept at age = 0~d
of the fit center coordinates ($\alpha_{\rm SNcenter}^i, \;
\delta_{\rm SNcenter}^i$) as a function of time.  The numbers in
parentheses are the standard errors in the last two digits.}

\tablenotetext{d}{The proper motion determined relative to M81\starcore.}

\tablenotetext{e}{The projected velocity determined relative to M81\starcore
(see text \S~\ref{errors}).}

\tablenotetext{f}{The projected velocity computed from the radial velocity
of the HI gas at the position of SN1993J by assuming circular motion
in the plane of the galaxy.}

\tablenotetext{g}{Vector subtraction of the projected velocity of the HI
gas at the position of SN1993J from the projected velocity of the geometric
center of the radio shell.}

\end{minipage}
\end{table}

\begin{deluxetable}{r@{ }l@{ }r r@{\hspace{0.3in}} c@{\hspace{0.6in}}  l l}
\tablecaption{Center Coordinates of SN1993J Relative to Explosion 
Center\label{cencoord}}
\tablewidth{0pt}
\tablehead{
\multicolumn{3}{c}{Date} &
\multicolumn{1}{l}{Age\tablenotemark{a}} &
\colhead{Epoch, $i$~~~} &
\colhead{$\Delta\alpha_i$~\tablenotemark{b}} &
\colhead{$\Delta\delta_i$~\tablenotemark{b}} \\
\colhead{} &
\colhead{} &
\colhead{} &
\multicolumn{1}{l}{(days)} & & 
\colhead{(mas)} &
\colhead{(mas)} 
}
\startdata
1993&May&17 &  50  & \phn 1 &\phs $0.005\pm0.040$  &\phs $0.040\pm0.040$ \\
1993&Jun&27 &  91  & \phn 2 &\phs 0.002   &\phs 0.010\\
1993&Sep&19 & 175  & \phn 3 &\phs 0.056   &\phs 0.013\\
1993&Nov& 6 & 223  & \phn 4 &\phs 0.042   &  $-$0.070\\
1993&Dec&17 & 264  & \phn 5 &   $-0.111$  &\phs 0.078\\
1994&Jan&28 & 306  & \phn 6 &   $-0.003$  &  $-$0.025\\
1994&Mar&15 & 352  & \phn 7 &\phs 0.008   &  $-$0.041\\
1994&Apr&22 & 390  & \phn 8 &   $-0.015$  &  $-$0.092\\
1994&Jun&22 & 451  & \phn 9 &   $-0.034$  &  $-$0.013\\
1994&Aug&30 & 520  &   10 &\phs 0.031   &\phs 0.023\\
1994&Oct&31 & 582  &   11 &   $-0.019$  &  $-$0.035\\
1994&Dec&23 & 635  &   12 &\phs 0.007   &  $-$0.027\\
1995&Feb&12 & 686  &   13 &   $-0.022$  &  $-$0.007\\
1995&May&11 & 774  &   14 &\phs 0.007   &\phs 0.035\\
1995&Aug&18 & 873  &   15 &   $-0.026$  &  $-$0.064\\
1995&Dec&19 & 996  &   16 &   $-0.044$  &  $-$0.026\\
1996&Apr& 8 &1107  &   17 &   $-0.054$  &  $-$0.031\\
1996&Sep& 1 &1253  &   18 &\phs 0.007   &\phs 0.020\\
1996&Dec&13 &1356  &   19 &   $-0.083$  &  $-$0.053\\
1997&Jun& 7 &1532  &   20 &   $-0.020$  &  $-$0.017\\
1997&Nov&15 &1693  &   21 &   $-0.014$  &  $-$0.071\\
1998&Jun& 3 &1893  &   22 &   $-0.025$  &  $-$0.070\\
1998&Dec& 7 &2080  &   23 &\phs 0.018   &  $-$0.112\\
2000&Feb&25 &2525  &   24 &   $-0.062$  &  $-$0.007\\
\enddata
\tablenotetext{a}{Time since shock breakout on 28.0 March 1993.}
\tablenotetext{b}{$\Delta\alpha_i, \; \Delta\delta_i$: The coordinates of
the geometric center of the fit model of the radio shell of SN1993J
($\alpha_{\rm SNcenter}^i, \; \delta_{\rm SNcenter}^i$) minus the
extrapolated coordinates of the center of explosion  ($\alpha_{\rm explosion}
= \Ra{9}{55}{24}{7747593}, \; \delta_{\rm explosion} =
\dec{69}{1}{13}{703188}$) for each epoch $i$.  The uncertainties are
standard errors. For more information, see text 
\S\S~\ref{datreds}\nolinebreak --\nolinebreak \ref{results}.}

\end{deluxetable}

\end{document}